\documentclass{aa}
\usepackage{epsfig}
\begin{document}

   \thesaurus{11     
              (11.01.1;  
               11.05.1;  
               11.05.2;  
               11.09.4;  
               09.16.1)}  
   \title{The chemical evolution of dynamically hot galaxies}


   \author{Michael G. Richer \inst{1}
          \thanks{Visiting Astronomer, Canada-France-Hawaii Telescope, operated
           by the National Research Council of Canada, le Centre National de la
           Recherche Scientifique de France, and the University of Hawaii}
          \and Marshall L. McCall \inst{2}
          \and Gra\. zyna Stasi\'nska \inst{3}$\ {}^{\star}$}

   \offprints{M. G. Richer}

   \institute{Instituto de Astronom\'{\i}a, UNAM,
              Apartado Postal 70-264, 04510 M\'exico, D. F., M\'exico\\
              email: richer@astroscu.unam.mx
              \and
              Dept. of Physics and Astronomy, York University,
              4700 Keele Street, Toronto, Ontario, Canada   M3J 1P3\\
              email: mccall@aries.phys.yorku.ca
              \and
              DAEC, Observatoire de Meudon,
              5 Place Jules Janssen, F-92195 Meudon Cedex, France\\
              email: grazyna@obspm.fr}

   \date{Received ? / Accepted ?}

   \maketitle

\begin{abstract}

We investigate the chemical properties of \object{M32}, the bulges
of \object{M31} and the \object{Milky Way}, and the dwarf
spheroidal galaxies \object{NGC 205}, \object{NGC 185},
\object{Sagittarius}, and \object{Fornax} using previously
published oxygen abundances for their planetary nebulae.  Our
principal result is that the mean stellar oxygen abundances for
all of these galaxies correlate very well with their mean velocity
dispersions.  This implies that the balance between energy input
from type II supernovae and the gravitational potential energy
controls how far chemical evolution proceeds in bulges,
ellipticals, and dwarf spheroidals.  It appears that chemical
evolution ceases once supernovae have injected sufficient energy
that a galactic wind develops.  All of the galaxies follow a
single relation between oxygen abundance and luminosity, but the
dwarf spheroidals have systematically higher [O/Fe] ratios than
the other galaxies. Consequently, dynamically hot galaxies do not
share a common star formation history nor need they share a common
chemical evolution, despite attaining similar mean stellar oxygen
abundances when forming similar stellar masses.  The oxygen
abundances support previous indications that the stars in higher
luminosity ellipticals and bulges were formed on a shorter time
scale than their counterparts in less luminous systems.  We
calibrate the Mg$_2$ index for dynamically hot galaxies as a
function of observed age and oxygen abundance using the mean
oxygen abundances from the planetary nebulae and the best age
estimates available for the stars in \object{M32} and the bulges
of \object{M31} and the \object{Milky Way}.


\end{abstract}

\section{Introduction}

At the present epoch, chemical evolution has largely ceased in
most dynamically hot galaxies (DHGs: ellipticals, bulges, and
dwarf spheroidals).  The majority of the stars in the majority of
DHGs are at least several billion years old, so it is likely that
we observe most DHGs in the chemical evolution state in which they
are destined to remain.  The properties of DHGs, however, reflect
the evolution that occurred earlier.  The existence of a
well-defined relation between metallicity and mass is a
fundamental clue concerning the evolution of DHGs (e.g., Bender et
al. \cite{Benderetal1993}).  The relation between the Mg$_2$ index
and the velocity dispersion, for instance, has long been
interpreted as evidence for galactic winds (e.g., Brocato et al.
\cite{Brocatoetal1990}). The velocity dispersion is a measure of
the galaxy's gravitational potential, and consequently is a probe
of its mass.  Type II supernovae are the principal source of
magnesium, so the Mg$_2$ index is a natural representative for the
energy injected by massive stars.  Larson (\cite{Larson1974})
first considered the effects of supernova-driven galactic winds
upon a star-forming elliptical galaxy:  As star formation
proceeds, supernovae explode, and part of their energy goes into
raising the internal energy of the interstellar medium.  If the
rate of energy input is higher than that at which it can be lost,
e.g., by radiative processes, the internal energy of the
interstellar medium will eventually exceed its gravitational
binding energy, and the interstellar medium will flow away in a
wind.

Progressing beyond this qualitative picture to a more quantitative
one has been difficult.  The difficulty arises largely because no
easily interpretable, quantitative measure of abundances has been
available for DHGs.  Traditionally, investigations of the
metallicities and chemical evolution of DHGs  have hinged upon
metallicity indicators derived from integrated light spectroscopy;
the Mg$_2$ index is the best known (see, e.g., Worthey et al.
\cite{Wortheyetal1994} for a definition). While the Mg$_2$ index
is an excellent tool for ranking metallicities in DHGs, it does
not directly yield an abundance, i.e., the number density of an
element relative to hydrogen. Through population syntheses, it is
possible to calibrate metallicity indicators in terms of the age
and metallicity of the underlying stellar population (e.g.,
Worthey \cite{Worthey1994}; Casuso et al. \cite{Casusoetal1996}),
but the utility of such calibrations is inevitably compromised due
to the difficulty of disentangling age and metallicity in
composite spectra (Worthey \cite{Worthey1994}). (By \lq\lq stellar
populations" we mean stars of a given age and metallicity,
recognizing that the metallicities of stars in a DHG may vary
considerably though their ages might not.)

Recently, CCD detectors coupled to efficient spectrographs on
4m-class telescopes have allowed direct measurements of oxygen
abundances for individual planetary nebulae in several nearby
DHGs.  Oxygen abundances are now available for planetary nebulae
in \object{M32}, the bulges of \object{M31} and the \object{Milky
Way}, and the dwarf spheroidals \object{NGC 205}, \object{NGC
185}, \object{Sagittarius}, and \object{Fornax}.  These
observations have considerably changed the complexion of the
abundance problem in DHGs.

In this paper, we focus specifically upon the energy balance and
temporal aspects of chemical evolution in DHGs, independent of
models.  In a following paper, we address issues such as the yield
of oxygen, the gas fractions when star formation ceased, and the
importance of gas flows during the period of star formation.  We
begin by examining how well the oxygen abundances in planetary
nebulae should trace those in the stars in DHGs (Section 2). Next,
we define our criteria for selecting planetary nebulae in DHGs and
outline the limitations of the mean oxygen abundances derived from
these samples (Section 3).  Then, we consider the implications of
the data for the chemical evolution of DHGs, paying particular
attention to the energy balance between supernovae and the
gravitational potential, and the time scale for star formation
(Section 4).  Following that, we present our calibration of the
Mg$_2$ index (Section 5).  Finally, we summarize our conclusions
(Section 6).

\section{Planetary nebulae as probes of stellar oxygen abundances}

The oxygen abundances in planetary nebulae offer several
advantages and new possibilities for studying the chemical
evolution of DHGs.  First, spectroscopic studies of planetary
nebulae yield oxygen abundances directly.  Second, oxygen is
produced almost exclusively by type II supernovae (e.g., Timmes et
al. \cite{Timmesetal1995}), so its evolution within a galaxy is
easily understood and modelled.  Third, the precursor stars to
planetary nebulae do not modify their initial oxygen abundances
significantly (e.g., Richer \cite{Richer1993}; Forrestini \&
Charbonnel \cite{ForestiniCharbonnel1997}). Fourth, the mean
oxygen abundance found for a planetary nebula population should be
a mass-weighted mean of the oxygen abundances in the stellar
populations from which the planetary nebula population arises.
Since the stellar populations comprising most of the mass in any
given DHG are old, they should have similar rates of both stellar
death and planetary nebula production (e.g., Renzini \& Buzzoni
\cite{RenziniBuzzoni1986}).  Consequently, the stellar populations
in DHGs should produce planetary nebulae in numbers proportional
to the mass of each stellar population.  Unless there are strong
gradients in the production of bright planetary nebulae among the
stellar populations in DHGs, the mean oxygen abundance for bright
planetary nebulae should be a mean of the stellar oxygen abundance
weighted according to the mass in different stellar populations.
In contrast, spectroscopic metallicity indicators weight stellar
populations according to their luminosity. Finally, a comparison
of the oxygen and iron abundances, when the latter exist, can be
used as a measure of the star formation time scale.  As iron is
produced in significant quantities by type Ia supernovae as well
as type II supernovae, its enrichment time scale is significantly
longer than that for oxygen, so the ratio of the two abundances
depends upon the history of star formation and is a reasonably
reliable measure of the star formation time scale (see Sects. 4.1
and 4.2).

While Richer (\cite{Richer1993}) showed that bright planetary
nebulae are good tracers of the interstellar medium abundances in
the \object{Magellanic Clouds}, doing so for DHGs is much more
difficult.  The \object{Milky Way} bulge is the only system in
which we can directly compare oxygen abundances in stars and
planetary nebulae.  McWilliam \& Rich (\cite{McWilliamRich1994})
found $[\mathrm{Fe}/\mathrm{H}] = -0.23 \pm 0.40$\,dex and
$[\mathrm{O}/\mathrm{Fe}] \sim 0$\,dex for the stars in the bulge.
Within errors, their oxygen abundance is identical to that quoted
for planetary nebulae in Table \ref{tab1}.  This agreement
suggests that the stars observed by McWilliam \& Rich
(\cite{McWilliamRich1994}) and the planetary nebulae selected by
Stasi\'nska et al. (\cite{Stasinskaetal1998}) probe the bulge's
stellar populations in the same way, and are thus directly
comparable.  (Throughout, we use logarithmic abundances relative
to their solar values, i.e., $[A] = \log A - \log A_\odot$, and
adopt the solar abundances of Anders \& Grevesse
(\cite{AndersGrevesse1989}).)

One worry is that McWilliam \& Rich (\cite{McWilliamRich1994})
derived large aluminum enhancements for the stars they observed.
This raises the spectre that whatever process enhances aluminum
and depletes oxygen in some globular cluster giants (e.g., Kraft
et al. \cite{Kraftetal1995}) may also be operating in bulge
giants.  Obtaining nitrogen and carbon abundances for these stars
would be extremely helpful in investigating this possibility, for
nitrogen is enhanced and carbon depleted in oxygen-depleted stars
in globular clusters.  The stars observed by McWilliam and Rich
(\cite{McWilliamRich1994}) have other abundance peculiarities,
notably the $\alpha$-elements Mg, Ca, Si, and Ti do not share a
common abundance enhancement relative to iron as is usually found.
This puzzling abundance pattern in bulge stars might simply
reflect a complex enrichment history, so one should consider the
possibility of oxygen depletion very carefully.  Briley et al.
(\cite{Brileyetal1996}) found sodium-enriched main sequence stars
in the globular cluster \object{47 Tuc}, suggesting that some of
these abundance anomalies might have a primordial origin (sodium
enrichment normally accompanies oxygen-depletion; e.g.,
Denissenkov et al. \cite{Denissenkovetal1998}). Both models and
observations hint that oxygen depletion is more efficient at low
metallicity, $[\mathrm{Fe}/\mathrm{H}] \leq -1$\,dex (e.g., Norris
\& Da Costa \cite{NorrisDaCosta1995}; Cavallo et al.
\cite{Cavalloetal1996}). Finally, abundance anomalies (and oxygen
depletion?) appear to be less frequent among metal-poor field
stars than among globular cluster giants (Langer et al.
\cite{Langeretal1992}; Pilachowski et al.
\cite{Pilachowskietal1996}).  It is therefore not clear whether
\object{Milky Way} bulge stars are oxygen-depleted, nor how the
oxygen abundances in planetary nebulae could be corrected to
account for oxygen depletion, since the depletion observed in
globular cluster stars varies among stars within a cluster and its
severity varies from cluster to cluster (e.g., Kraft et al.
\cite{Kraftetal1995}).

\begin{table*}
\caption[]{Elemental abundances$^{a}$ and galactic properties for
our sample of DHGs} \label{tab1}
\[
\begin{tabular}{lcccccccl}
\hline \noalign{\smallskip} Galaxy & N$_{\mathrm{PNe}}$ & mean
[O/H] & max [O/H] & mean [Fe/H] & M$_B$ & $\sigma_m$ & Mg$_2$ &
Data Sources$^{b}$ \\
       &                    & (dex)      & (dex)     & (dex)       &
(mag) & (km/s)     & (mag)  &              \\
\noalign{\smallskip}
\hline
\noalign{\smallskip}
MW bulge           & 32 & $-0.30 \pm 0.27$ & 0.27 & $-0.23 \pm 0.40$ &
-19.5 & 125 & 0.25 & A,F,M,N,U \\
\object{M31} bulge & 21 & $-0.26 \pm 0.27$ & 0.18 &         &
-19.6 & 166 & 0.34 & B,M,P,V   \\
\object{M32}       & 5  & $-0.61 \pm 0.20$ & -0.43& $-0.25 \pm 0.30$ &
-15.6 & 60  & 0.20 & B,G,M,S,V \\
\object{NGC 205}   & 8  & -0.40            &      & $-0.85 \pm 0.50$ &
-15.9 & 50  & 0.10 & C,H,C,Q,W \\
\object{NGC 185}   & 4  & -0.78            &      & $-1.23 \pm 0.30$ &
-14.6 & 25  & 0.08 & C,K,C,T,W \\
\object{Sagittarius} & 2 & -0.93           &      & $-1.10 \pm 0.30$ &
-13   & 11.4 &     & D,J,J,J   \\
\object{Fornax}    & 1  & -0.95            &      & $-1.34 \pm 0.40$ &
-11.7 & 9.6 & 0.067 & E,L,C,R,W \\
\noalign{\smallskip}
\hline
\end{tabular}
\]
\begin{list}{}{}
\item[$^{\mathrm{a}}$] Throughout, we use the notation
$[A] = \log A - \log A_\odot$ and the Anders \& Grevesse
(\cite{AndersGrevesse1989}) solar abundances.
\item[$^{\mathrm{b}}$] The order of the data sources is [O/H], [Fe/H],
M$_B$, $\sigma_m$, and Mg$_2$.
\item[] Data sources: (A) Stasi\'nska et al. \cite{Stasinskaetal1998};
(B) Richer et al. \cite{Richeretal1998}; (C) Richer \& McCall
\cite{RicherMcCall1995}; (D) Walsh et al. \cite{Walshetal1997};
(E) Maran et al. \cite{Maranetal1984}; (F) McWilliam \& Rich
\cite{McWilliamRich1994}; (G) Grillmair et al.
\cite{Grillmairetal1996}; (H) Mould et al. \cite{Mouldetal1984};
(J) Ibata et al. \cite{Ibataetal1997}; (K) Lee et al.
\cite{Leeetal1993}; (L) Beauchamp et al. \cite{Beauchampetal1995};
(M) McCall \cite{McCall1998}; (N) Sellgren et al.
\cite{Sellgrenetal1990}; (P) Kormendy \cite{Kormendy1988}; (Q)
average of Held et al. \cite{Heldetal1990}, Carter \& Sadler
\cite{CarterSadler1990}, and Bender et al. \cite{Benderetal1991};
(R) Mateo \cite{Mateo1997}; (S) Kormendy \cite{Kormendy1987}; (T)
average of Held et al. \cite{Heldetal1992} and Bender et al.
\cite{Benderetal1991}; (U) Whitford \cite{Whitford1978}; (V)
Worthey et al. \cite{Wortheyetal1992}; (W) Bender et al.
\cite{Benderetal1993}
\end{list}
\end{table*}

The bright planetary nebulae in the \object{Milky Way} bulge,
however, do not have the chemical signatures of the
oxygen-depleted stars in globular clusters.  The bright planetary
nebulae in the \object{Milky Way} bulge have [N/O] ratios
identical to those found in bright planetary nebulae in the
Magellanic Clouds (Stasi\'nska et al. \cite{Stasinskaetal1998}),
and the latter have not significantly depleted their initial store
of oxygen (Richer \cite{Richer1993}). In fact, no planetary nebula
sample for any of the galaxies we consider here has the high [N/O]
ratios observed in oxygen-depleted giants (e.g.,
$[\mathrm{N}/\mathrm{O}] > 1$\,dex; Kraft et al.
\cite{Kraftetal1995}) or the even higher ratios allowed by models
(e.g., Denissenkov et al. \cite{Denissenkovetal1998}). Similarly,
these same models predict much larger [Ne/O] ratios for
oxygen-depleted stars than are observed, on average, in the
planetary nebula sample in the \object{Milky Way} bulge (or
anywhere; Stasi\'nska et al. \cite{Stasinskaetal1998}). In short,
the bright planetary nebulae in the \object{Milky Way} bulge
appear to have normal abundance ratios, and so they presumably
probe the oxygen abundances when their progenitor stars formed.

A second worry is that the number of bright planetary nebulae per
unit luminosity in external galaxies decreases in more luminous or
redder galaxies (e.g., Hui et al. \cite{Huietal1993}). These
trends are of unknown origin, effects due to both age and
metallicity of the progenitor stellar populations have been
suggested, and they could have an impact when comparing samples of
planetary nebulae in galaxies of very different luminosities. If a
metallicity effect is involved, planetary nebulae will sample the
mean oxygen abundance in stars in a systematically different way
in galaxies of different luminosity.  If an age effect is
involved, it will probably only affect the mean oxygen abundance
obtained from planetary nebulae if a galaxy contains stellar
populations of such different ages that the stellar death rates
from these populations are markedly different.

Neither of these worries should invalidate our approach. Although
it is unclear whether the stars in the \object{Milky Way} bulge do
or should have modified abundances, the planetary nebulae have
normal abundances.  Likewise, the abundances for the planetary
nebulae in \object{M31}, \object{M32}, and \object{NGC 205} (Table
\ref{tab1} and Stasi\'nska et al. \cite{Stasinskaetal1998}) appear
to be unmodified, consistent with the arguments presented above.
For the planetary nebulae in \object{Sagittarius} and
\object{Fornax}, both [N/O] and [Ne/O] are normal (Walsh et al.
\cite{Walshetal1997}; Maran et al. \cite{Maranetal1984}), so they
presumably likewise measure the initial oxygen abundances in the
stars from which they descend. By interpolation, we have no reason
to believe that the planetary nebulae in \object{NGC 185} should
have oxygen abundances that have been significantly modified by
the evolution of their progenitors.  In the very worst case, the
stars in the \object{Milky Way} bulge could be oxygen-depleted,
and the planetary nebulae then under-estimate the mean initial
abundance for the stars, since the planetary nebula abundances
then agree with those in oxygen-depleted stars. Such an error is
in the sense of being conservative, and even such a lower limit is
very useful for our purposes.  At any rate, should any pathology
affect the relationship between oxygen abundances in stars and
planetary nebulae in the \object{Milky Way}, we assume that the
same effect occurs in all of the DHGs included here.  Finally, the
trends in planetary nebula production in different galaxies do not
appear to seriously affect our results, for our results agree with
completely independent evidence concerning the evolution and star
formation history of DHGs (Section 4.2).

\section{Mean stellar oxygen abundances}

Spectroscopic observations exist of individual planetary nebulae
in all of the galaxies listed in Table \ref{tab1}.  To uniformly
sample the stellar populations, we restrict our attention to the
brightest planetary nebulae in each galaxy, specifically those
within 2\,mag of the peak of the planetary nebula luminosity
function.  For these bright planetary nebulae, Table \ref{tab1}
quotes the mean and the standard deviation of $12 + \log
(\mathrm{O}/\mathrm{H})$ as well as the number of objects included
in these calculations.  We adopt these mean oxygen abundances for
the bright planetary nebulae as the mean oxygen abundances for the
stars.  The source of the original spectroscopic data is found in
the last column.

It is important to note that the manner in which the mean oxygen
abundance was calculated varied from galaxy to galaxy.  For the
\object{Milky Way}, electron temperatures exist for every
planetary nebula in the sample save two. Apart from these two
exceptions, for which we could only derive lower limits to the
oxygen abundance, we were able to determine the actual oxygen
abundances and obtain a mean value directly.  For \object{M32} and
the bulge of \object{M31}, electron temperatures and the
corresponding oxygen abundances could be measured for only half of
the planetary nebulae (Richer et al. \cite{Richeretal1998}). For
the other half, only temperature-based lower limits to the oxygen
abundance exist. Thus, the tabulated mean abundances for
\object{M32} and the bulge of \object{M31} are lower limits to the
true mean values, but are likely within $\sim 0.1$\,dex of their
true values (Stasi\'nska et al. \cite{Stasinskaetal1998}; Richer
et al. \cite{Richeretal1998}).

For the planetary nebulae in \object{NGC 185} and \object{NGC
205}, only empirical lower limits to the oxygen abundances exist
(Richer \& McCall \cite{RicherMcCall1995}). We re-calculated the
limits because the original calibration did not allow for the
large values of [\ion{O}{iii}]$\lambda 5007 /\mathrm{H}\beta$
observed in planetary nebulae in \object{M31}.  On average, the
re-calibrated abundances for the planetary nebulae in \object{NGC
185} and \object{NGC 205} are slightly lower than the previous
values.  We used these limits for individual planetary nebulae to
estimate the mean oxygen abundance for the planetary nebula
population following the prescription given in Richer \& McCall
(\cite{RicherMcCall1995}). Various tests, based upon the planetary
nebulae in the bulges of \object{M31} and the \object{Milky Way}
and in the Magellanic Clouds, indicate that our estimates of the
mean oxygen abundances in \object{NGC 185} and \object{NGC 205}
should be within $\sim 0.2$\,dex of the actual mean values.

The oxygen abundances for the planetary nebulae in \object{Fornax}
and \object{Sagittarius} are known accurately, but \object{Fornax}
and \object{Sagittarius} have only one and two planetary nebulae,
respectively. Had we assumed that the oxygen abundances in their
planetary nebulae were equal to those in their stars, we would
obtain [O/Fe] ratios of 0.79\,dex and 0.47\,dex in \object{Fornax}
and \object{Sagittarius}, respectively.  Both values, but
particularly that for \object{Fornax}, are larger than the value
of 0.3 to 0.4\,dex observed in the \object{Milky Way} halo (e.g.,
Wheeler et al. \cite{Wheeleretal1989}), which presumably reflects
the maximum [O/Fe] ratio attainable (via pure type II supernova
enrichment). Therefore, we estimated mean stellar oxygen
abundances in \object{Fornax} and \object{Sagittarius} by
subtracting the dispersion in stellar iron abundances from the
mean oxygen abundances found in their planetary nebulae.

For all of the galaxies we consider, save \object{M31}, there exist
observations of individual stars from which mean iron abundances
have been derived.  In the bulge of the \object{Milky Way}, these results
come from medium-resolution spectroscopy, while in the other
galaxies they come from photometry.  Table \ref{tab1} lists the mean
stellar iron abundances, the dispersion about the mean, and the
original data sources for each galaxy (in the last column).
Table \ref{tab1} also lists absolute blue magnitudes, velocity
dispersions, and measured values of the Mg$_2$ index.  For \object{Fornax}
and \object{Sagittarius}, the velocity dispersion samples the entire
galaxy out to beyond one effective radius.  For the other
galaxies, the velocity dispersion samples the stellar motions
inside one effective radius, but excluding any cusps due to
central mass concentrations, e.g., as seen in \object{M31} and
\object{M32}.  The
Mg$_2$ index values are typically those for the nuclei.  The sources
for all of these data are also found in the last column of Table
\ref{tab1}.

\section{Chemical evolution of DHGs}

\subsection{The role of the gravitational potential}

\begin{figure*}
\vspace{-1.7cm}
\epsfig{file=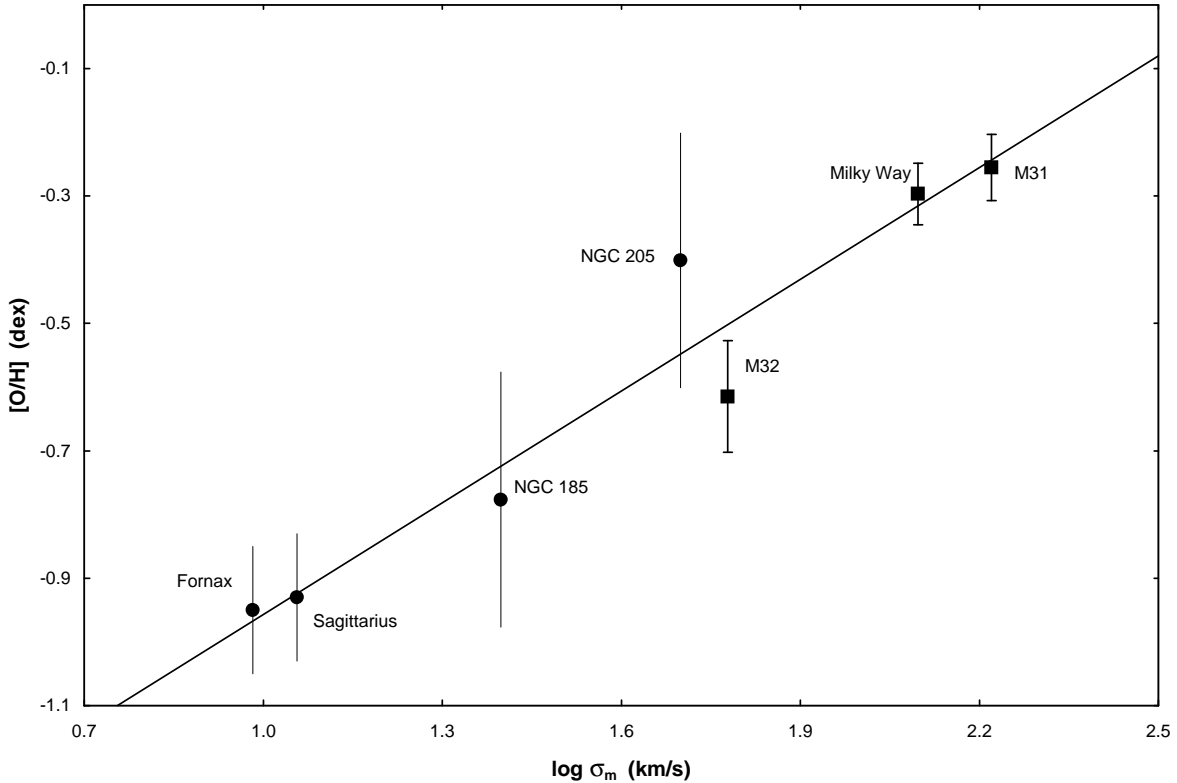,height=18cm,clip=true,angle=270}
\vspace{-1.2cm} \caption{ The oxygen abundances in DHGs are
correlated with their mean velocity dispersions.  Since the oxygen
abundance is a measure of the energy input from type II supernovae
and the velocity dispersion is a measure of the gravitational
potential, the correlation observed arises naturally if energy
input from supernovae is the agent that eventually stops chemical
evolution in DHGs.  The error bars denote the uncertainty in the
mean [O/H]: For the dwarf spheroidals, this uncertainty was
estimated empirically, but for \object{M32} and the bulges of
\object{M31} and the \object{Milky Way} the uncertainty is the
standard error in the mean. } \label{fig1}
\end{figure*}

The mean oxygen abundances attained in \object{M32}, and the
bulges of \object{M31} and the \object{Milky Way} are not
high. Even the maximum oxygen abundance observed is only about
twice the solar value. Since these values are modest, they require
no exotic stellar initial mass function (IMF) for their
production; they are easily achievable with a standard Salpeter
(\cite{Salpeter1955}) IMF (e.g., see Richer et al.
\cite{Richeretal1997} for the case of closed box models).

Intuitively, galaxies with deeper gravitational potential wells
should retain supernova ejecta more efficiently than galaxies with
shallower potentials, given broadly similar supernova rates. In
Fig. \ref{fig1}, we test this idea directly, plotting the mean
oxygen abundance as a function of the mean velocity dispersion for
the DHGs in Table \ref{tab1}. The correlation in Fig. \ref{fig1}
is remarkably tight, and suggests that a common mechanism controls
chemical evolution in all of the galaxies.  Oxygen is a product of
type II supernovae, so the oxygen abundance is a measure of the
energy input from type II supernovae.  The velocity dispersion is
related to the depth of the gravitational potential through the
virial theorem.  Thus, an [O/H]$- \sigma_m$ correlation arises
naturally if chemical evolution proceeds until the energy input
from supernovae has raised the thermal energy of the interstellar
medium beyond its gravitational binding energy, instigating the
development of a galactic wind (e.g., Larson \cite{Larson1974}).
Chemical evolution (and star formation) would cease with the loss
of the remaining gas.

There is some indication that the dwarf spheroidals define a
relation offset from that defined by \object{M32} and the bulges
of \object{M31} and the \object{Milky Way}.  This offset depends
strongly upon the positions of \object{M32} and \object{NGC 205},
the latter in particular, and to ascertain its reality would
require deeper spectroscopy of the planetary nebulae in
\object{NGC 205} and \object{NGC 185} to obtain direct estimates
of their mean oxygen abundances.  Even if the dwarf spheroidals
are offset from the other DHGs in Fig. \ref{fig1}, the conclusion
that galactic winds terminated chemical evolution in DHGs would
remain intact.  For example, an offset could arise if there were
systematic differences in gas flows prior to initiation of the
wind, say, as a result of differences in mass distributions.  What
matters is that the oxygen abundance and velocity dispersion are
correlated within each group.

Figure \ref{fig1} implies a tight connection between energy input
from type II supernovae and the gravitational potential.  However,
if energy input from supernovae is responsible for halting
chemical evolution, we ought to consider the energy contribution
from type Ia supernovae as well.  We can use the oxygen and iron
abundances for the stars in these galaxies (save M31) to calculate
the fraction of the iron contributed by type Ia supernovae.  Then,
using the oxygen and iron yields of type Ia and II supernovae, we
can calculate the fraction of all supernovae that were of type Ia.

\begin{figure*}
\vspace{-1.8cm}
\epsfig{file=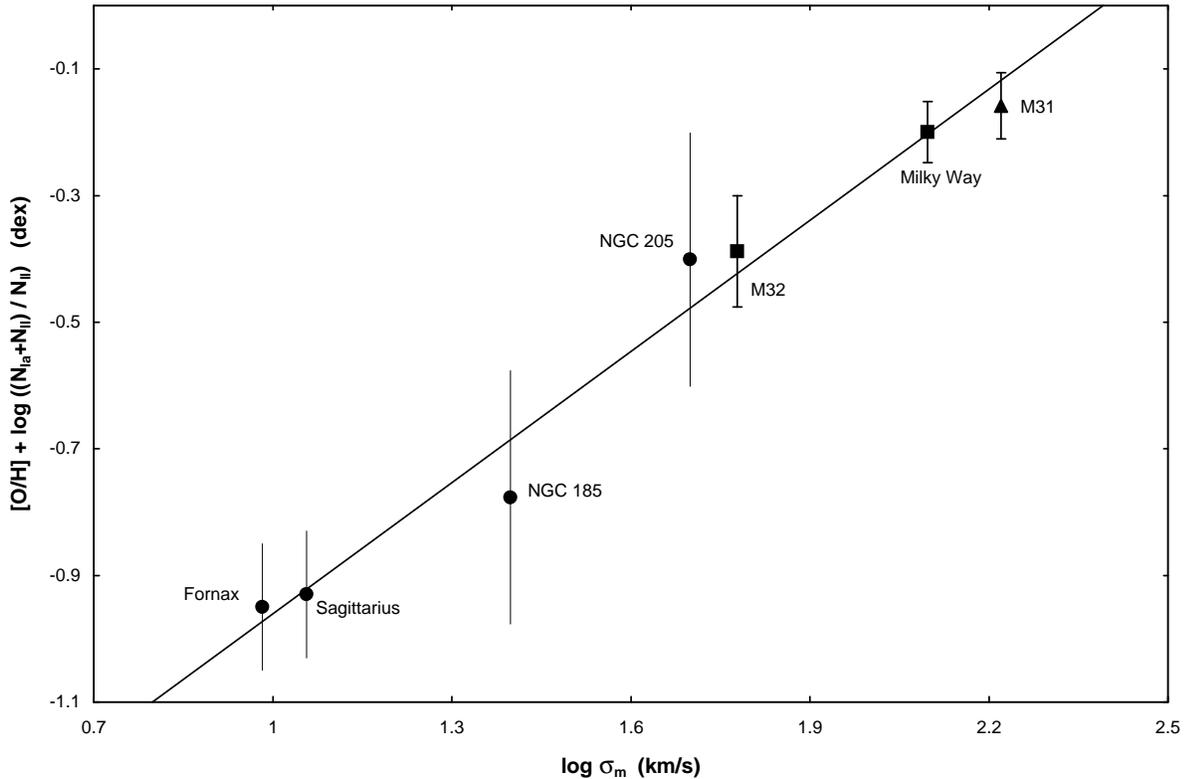,height=18cm,clip=true,angle=270}
\vspace{-1.2cm} \caption{ If we correct the oxygen abundances to
account for the energy injected by type Ia supernovae, the
correlation with the mean velocity dispersion improves.  The
required correction is obtained by multiplying the measured oxygen
abundances by the ratio of the total number of supernovae to the
number of type II supernovae, thereby accounting for the energy
input from type Ia supernovae (see text for details).  The
modified oxygen abundances are proportional to the total energy
input from all supernovae per hydrogen atom locked into stars. The
excellent correlation strengthens the conclusion that galactic
winds terminate chemical evolution in all DHGs.  The bulge of
\object{M31} is plotted with a different symbol since its [O/Fe]
ratio was assumed identical to that for the bulge of the
\object{Milky Way}.  The line is a fit to all of the points.  The
error bars denote the uncertainty in the mean [O/H] (see Fig.
\ref{fig1}). } \label{fig2}
\end{figure*}

We estimate the iron contribution from type Ia supernovae by
comparing the [O/Fe] ratios in galaxies where type Ia supernovae
had, respectively, significant and negligible effects.  For
galaxies in which type Ia supernovae contributed significantly to
the iron abundance, we can express the [O/Fe] ratio as
\begin{eqnarray}
[\mathrm{O}/\mathrm{Fe}] & = & \log(\mathrm{O}/\mathrm{Fe}) -
                                \log(\mathrm{O}/\mathrm{Fe})_\odot
                                \nonumber \\
   & = & \log\left(\left(\frac{\mathrm{O}}{\mathrm{Fe}}\right)_{II}
           \frac{1}{(1+f_I)}\right)
           - \log(\mathrm{O}/\mathrm{Fe})_\odot \label{eqn1}
\end{eqnarray}
where
$(\mathrm{O}/\mathrm{Fe})_{II}$
is the ratio of the oxygen to iron abundance by number
produced by type II supernovae, and
$f_I = M(\mathrm{Fe})_I/M(\mathrm{Fe})_{II}$
is the mass of iron
contributed by all type Ia supernovae relative to that
contributed by all type II supernovae.  For galaxies in which
only type II supernovae contributed to the iron abundance, the
[O/Fe] ratio,
$[\mathrm{O}/\mathrm{Fe}]_{II}$, is obviously
\begin{equation}
[\mathrm{O}/\mathrm{Fe}]_{II} = \log\left(\frac{\mathrm{O}}{\mathrm{Fe}}
\right)_{II} - \log(\mathrm{O}/\mathrm{Fe})_\odot\,.
\label{eqn2}
\end{equation}
Subtracting Eq. \ref{eqn2} from Eq. \ref{eqn1} yields
\begin{eqnarray*}
\Delta[\mathrm{O}/\mathrm{Fe}] = [\mathrm{O}/\mathrm{Fe}] -
[\mathrm{O}/\mathrm{Fe}]_{II} = \log\frac{1}{1+f_I} \nonumber
\end{eqnarray*}
\begin{equation}
\mathrm{or\ \ } f_I = 10^{-\Delta[\mathrm{O}/\mathrm{Fe}]} - 1\,.
\label{eqn3}
\end{equation}
Finally, the relative number of type Ia and II supernovae is
given by
\begin{equation}
\frac{N_{SNI}}{N_{SNII}} = f_I \frac{Y_{SNII}}{Y_{SNI}}
\label{eqn4}
\end{equation}
where $Y_{SNI}$ and $Y_{SNII}$ are the average masses of iron, in
solar masses, produced by individual type Ia and type II
supernovae, respectively. We adopt $[\mathrm{O}/\mathrm{Fe}]_{II}
= 0.30$\,dex, as found for halo stars in the \object{Milky Way}
(e.g., Wheeler et al. \cite{Wheeleretal1989}).   Within errors,
this value coincides with the [O/Fe] values observed in the dwarf
spheroidal galaxies, so it seems reasonable to assume that type II
supernovae completely dominated their iron production.  In
\object{M32} and the bulge of the \object{Milky Way}, Eq.
\ref{eqn3} and the oxygen and iron abundances in Table \ref{tab1}
indicate that type Ia supernovae provided 3.6 times as much iron
as type II supernovae in \object{M32}, and 1.3 times as much iron
as type II supernovae in the bulge of the \object{Milky Way}.

For type Ia supernovae, we adopt an iron yield of 0.63 $M_\odot$
(Thielemann et al. \cite{Thielemannetal1986}). For type II
supernovae, we prefer to estimate the typical yield of iron given
the predicted yield of oxygen and the [O/Fe] ratio observed in
halo stars, for the oxygen yields are less model-dependent than
the iron yields (cf. Woosley \& Weaver \cite{WoosleyWeaver1995}
and Thielemann et al. \cite{Thielemannetal1996}). Convolving the
oxygen yields from Woosley \& Weaver (\cite{WoosleyWeaver1995};
their \lq\lq A" sequences for $Z=Z_\odot$, $0.1\,Z_\odot$, and
$0.01\,Z_\odot$) with a Salpeter (\cite{Salpeter1955}) IMF, we
find that a typical type II supernova produces 1.8\,$M_\odot$ of
oxygen (we would obtain a similar result using the models of
Thielemann et al. \cite{Thielemannetal1996}). For
$[\mathrm{O}/\mathrm{Fe}]_{II} = 0.30$\,dex and the solar oxygen
and iron abundances (Anders \& Grevesse
\cite{AndersGrevesse1989}), we deduce that a typical type II
supernova produces 0.12\,$M_\odot$ of iron.  By Eq. \ref{eqn4},
the number of type Ia supernovae per type II supernova was 0.68 in
\object{M32}, 0.25 in the bulge of the \object{Milky Way}, and 0
in the dwarf spheroidals.  Thus, the ratio of all supernovae to
type II supernovae was 1.68, 1.25, and 1.0 in \object{M32}, the
bulge of the \object{Milky Way}, and the dwarf spheroidals,
respectively.

If we assume that type Ia and II supernovae inject comparable
quantities of energy into the interstellar medium (e.g., Woosley
\& Weaver \cite{WoosleyWeaver1986}), we can convert the oxygen
abundances in Fig. \ref{fig1} into measures of the total energy
injected by all supernovae.  To do so, we multiply the oxygen
abundances in \object{M32}, the bulge of the \object{Milky Way},
and the dwarf spheroidals by 1.68, 1.25, and 1.0, respectively.
Effectively, these modified oxygen abundances are the oxygen
abundances these galaxies would have if all of their supernovae
had been of type II.

These modified oxygen abundances, now a measure of the energy
injected by all supernovae, are plotted as a function of the mean
velocity dispersion in Fig. \ref{fig2}. The oxygen abundance for
the bulge of \object{M31} (plotted with a different symbol) was
corrected assuming its [O/Fe] ratio is identical to that for the
\object{Milky Way} bulge. The line is a least squares fit to all
of the points.  Figure \ref{fig2} shows that the total energy
injected by all supernovae is extremely well correlated with the
velocity dispersion.  This further supports the contention that
energy injection by supernovae is the mechanism that leads to
termination of the chemical evolution of DHGs.

None of the foregoing excludes the formation of DHGs through
mergers, but these findings do constrain mergers somewhat.  To
achieve a correlation between oxygen abundance and velocity
dispersion (Fig. \ref{fig1}), the majority of the stars must have
been formed either as a result of mergers, or in a potential
comparable to that in which they now reside.  In other words,
mergers would have had to have occurred between systems with much
gas or between systems of similar size.  None of these arguments
prevent minor mergers from occurring up to the present.

\subsection{The star formation time scale}

The [O/Fe] ratios for the dwarf spheroidals are systematically
higher than those for \object{M32} and the bulge of the
\object{Milky Way}. It is also likely that they are higher than in
the bulge of \object{M31}, since the bulge of \object{M31} is
believed to have a higher nitrogen abundance than the bulge of the
\object{Milky Way} (Worthey \cite{Worthey1996}) and nitrogen, like
iron, is enriched on long time scales (e.g., Timmes et al.
\cite{Timmesetal1995}). A related result is Jablonka et al.'s
(\cite{Jablonkaetal1996}) finding that [Mg/Fe] in bulges
correlated with luminosity.  Similarly, J{\o}rgensen
(\cite{Jorgensen1997}) found that [Mg/Fe] correlated with velocity
dispersion in elliptical and S0 galaxies in clusters.  Since both
oxygen and magnesium are $\alpha$-capture elements produced by
type II supernovae (formed by successive captures of
$\alpha$-particles on ${}^{12}$C seeds: O, Ne, Mg, Si, S, Ar, Ca,
and Ti), these [O/Fe] and [Mg/Fe] ratios have important
implications for the formation time scales of bulges and
ellipticals.

The time scale for star formation probably has the strongest
influence in determining [O/Fe] or [Mg/Fe]. Shortening the period
of star formation, flattening the IMF, or selectively losing type
Ia supernova ejecta will all increase [O/Fe] or [Mg/Fe] (e.g.,
Worthey et al. \cite{Wortheyetal1992}). Shortening the duration of
star formation enhances [O/Fe] since there is a fixed time lag
before the production of iron from type Ia supernovae. Selective
loss of type Ia ejecta enhances [O/Fe] since this iron production
is never incorporated into stars.  This might occur if forming and
previously-formed stars have different spatial distributions, an
effect that is more likely to occur if star formation lasts a long
time and the star-forming gas settles through dissipation.
Flattening the IMF enhances oxygen production by converting a
larger fraction of the baryonic mass into the high mass stars that
become type II supernovae. Recent studies, however, indicate that
the IMF is independent of metallicity, galactic environment, star
cluster density, and stellar mass range (e.g., Hill et al.
\cite{Hilletal1994}; Massey et al. \cite{Masseyetal1995}; Hunter
et al. \cite{Hunteretal1997}; Chabrier \& Mera
\cite{ChabrierMera1997}; Grillmair et al.
\cite{Grillmairetal1998}). So, it is unlikely that the IMF affects
[O/Fe] or [Mg/Fe].

\begin{figure}
\vspace{-0.8cm}
\epsfig{file=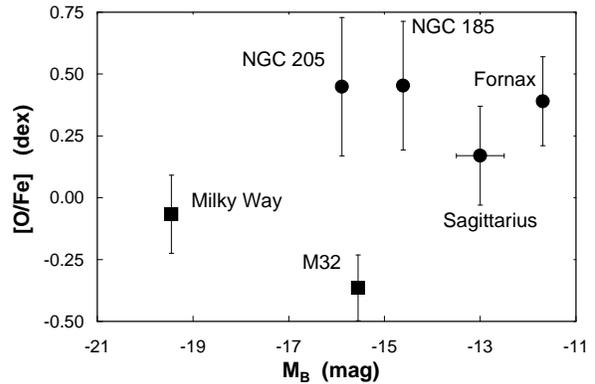,height=9cm,clip=true,angle=270}
\vspace{-0.6cm} \caption{ The relationship between [O/Fe] and
luminosity: This is a test of whether the IMF influences the value
of [O/Fe].  For a flatter IMF, the luminosity is lower and the
[O/Fe] ratio is higher when star formation stops.  \object{M32}
and \object{NGC 205} clearly contradict these expectations:  If
\object{M32} achieved its low [O/Fe] ratio via a steep IMF, it
should have a higher luminosity than \object{NGC 205}, for
\object{M32} has the deeper gravitational potential.  We therefore
conclude that the [O/Fe] ratio is set mainly by the time scale for
star formation. } \label{fig3}
\end{figure}

\begin{figure*}
\vspace{-1.8cm}
\epsfig{file=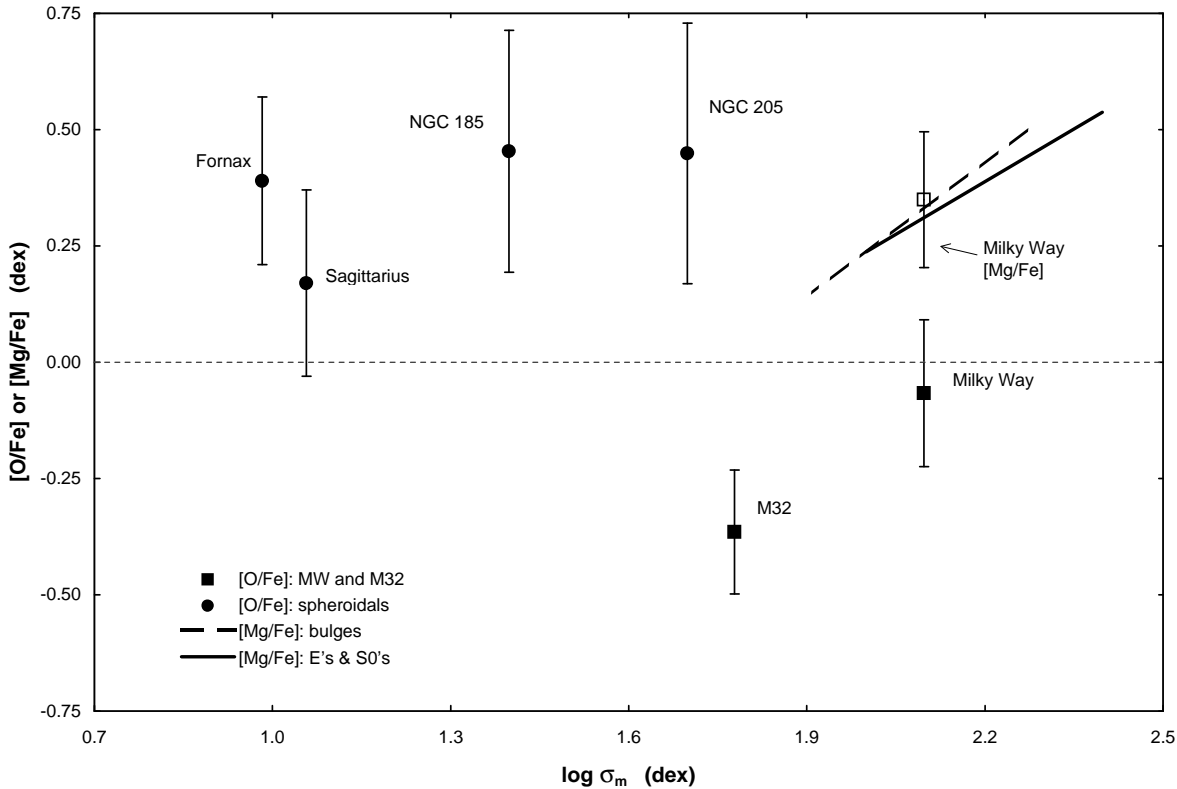,height=18cm,clip=true,angle=270}
\vspace{-1.2cm} \caption{ [O/Fe] and [Mg/Fe] in DHGs:  The [O/Fe]
ratios derived from Table \ref{tab1} for the DHGs in our sample
are plotted along with [Mg/Fe] relations for bulges and for
elliptical and S0 galaxies (Jablonka et al. 1996; J{\o}rgensen
1997).  The zero point for the relation for ellipticals and S0's
was undefined, so this relation has been arbitrarily adjusted to
match that for bulges at a velocity dispersion of 100 km/s.  The
horizontal line denotes the solar [O/Fe] and [Mg/Fe] ratios.  The
mean [Mg/Fe] ratio and $\pm 1\,\sigma$ dispersion observed in the
bulge of the \object{Milky Way} are shown for comparison
(McWilliam \& Rich 1994).  The error bars for the [O/Fe] values
account for the uncertainties in both [O/H] and [Fe/H]. Although
the [O/Fe] ratios for \object{M32} and the bulge of the
\object{Milky Way} track the [Mg/Fe] relations, the [O/Fe] ratios
for the dwarf spheroidals do not, and are instead consistent with
the value expected for enrichment purely from type II supernovae.
A correlation of either [O/Fe] or [Mg/Fe] with velocity dispersion
implies that the gravitational potential fixes the time scale for
star formation.  The [Mg/Fe] relations and the oxygen abundances
in \object{M32} and the bulge of the \object{Milky Way} require
that the star formation time scale be shorter in more massive
ellipticals and bulges. } \label{fig4}
\end{figure*}

The data presented here also favours a constant IMF in the context
of supernova-driven winds. Per unit mass of stars formed, a
flatter IMF injects more energy and oxygen from type II supernovae
into the interstellar medium, but produces fewer long-lived stars.
For a flatter IMF, a galaxy will have a lower luminosity and
higher [O/Fe] when star formation ceases.  This ought to be a
sensitive test since both oxygen production and the fraction of
long-lived stars depend sensitively upon the IMF slope for slopes
near the Salpeter (\cite{Salpeter1955}) value (K\"oppen \& Arimoto
\cite{KoppenArimoto1991}). Figure \ref{fig3} plots the [O/Fe]
ratio as a function of absolute blue luminosity for the galaxies
in Table \ref{tab1}.  Compare \object{M32} and \object{NGC 205}.
If we attempt to explain the low value of [O/Fe] in \object{M32}
as a consequence of a steep IMF, we face a quandary concerning its
luminosity.  \object{M32} should then have a much higher
luminosity than \object{NGC 205}, not only on account of its
supposedly steeper IMF, but also because it has a slightly deeper
gravitational potential, and much more star formation would then
have been required to accumulate the necessary energy from
supernovae to initiate a galactic wind. Yet, \object{M32} has the
lower luminosity. \footnote{ One might argue that \object{M32} has
a low luminosity on account of tidal stripping by \object{M31},
but its structural parameters are perfectly typical for an
elliptical of its luminosity (Kormendy \cite{Kormendy1985}; Bender
et al. \cite{Benderetal1993}). } That \object{M32} formed the
majority of its stars on a longer time scale and that type Ia
supernovae provided the additional energy required to initiate a
galactic wind is a far simpler explanation. Henceforth, we shall
assume that the star formation time scale governs [O/Fe] or
[Mg/Fe].

\begin{figure*}
\vspace{-1.6cm}
\epsfig{file=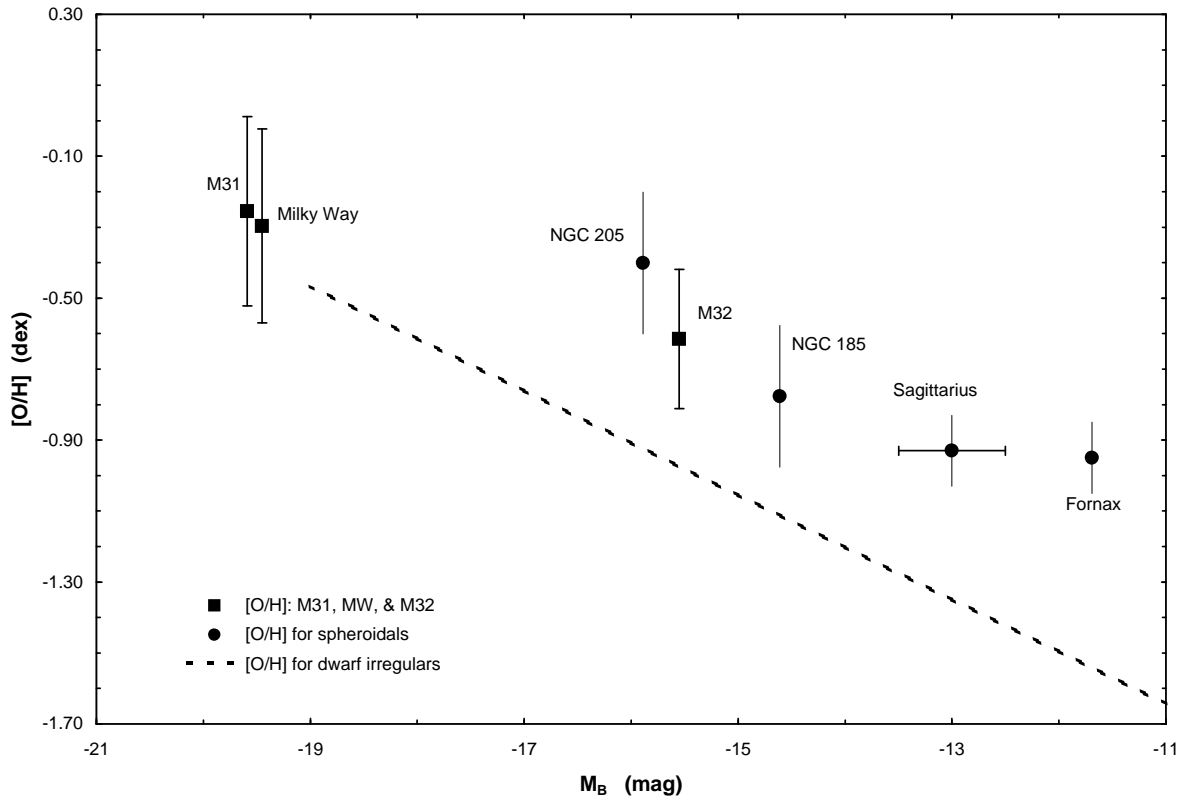,height=18cm,clip=true,angle=270}
\vspace{-1.2cm} \caption{ Oxygen abundance-luminosity relations
for galaxies: Oxygen abundances in DHGs are compared with those in
dwarf irregular galaxies (Richer \& McCall
\cite{RicherMcCall1995}). The luminosities are compared through
absolute blue magnitudes.  For \object{M32} and the bulges of
\object{M31} and the \object{Milky Way}, the error bars on the
oxygen abundances denote the standard deviations of the abundance
distributions.  For the dwarf spheroidals, the error bars denote
the uncertainty in the mean oxygen abundance.  We assume a
0.5\,mag uncertainty for the luminosity of \object{Sagittarius},
since it is derived rather than measured (Ibata et al.
\cite{Ibataetal1997}). Note that the oxygen abundance-luminosity
relation for DHGs pertains to the mean abundance for the stars,
whereas that for the dwarf irregulars pertains to the interstellar
medium, i.e., the maximum abundance for the stars.  Somewhat
surprisingly, all DHGs, even dwarf spheroidals, follow a single
[O/H]-M$_B$ relation. } \label{fig5}
\end{figure*}

In Fig. \ref{fig4}, we plot the [O/Fe] ratios for the DHGs in
Table \ref{tab1} as a function of the mean velocity dispersion. In
addition, we plot McWilliam \& Rich's (\cite{McWilliamRich1994})
[Mg/Fe] measurement for the \object{Milky Way} bulge and the
[Mg/Fe] relations found by Jablonka et al.
(\cite{Jablonkaetal1996}) and J{\o}rgensen (\cite{Jorgensen1997}).
Jablonka et al.'s (\cite{Jablonkaetal1996}) relation is shown as a
long-dashed line in Fig. \ref{fig4}.  We converted from luminosity
to velocity dispersion using their measured values of the Mg$_2$
index and the relation between the Mg$_2$ index and velocity
dispersion from Bender et al. (\cite{Benderetal1993}). The
relation found by J{\o}rgensen (\cite{Jorgensen1997}) is shown as
a solid line. Since J{\o}rgensen's (\cite{Jorgensen1997}) relation
was purely differential, we arbitrarily set its zero point so that
it matched Jablonka et al.'s (\cite{Jablonkaetal1996}) relation at
a velocity dispersion of 100 km/s. The dashed horizontal line
denotes the solar [O/Fe] and [Mg/Fe] values.

The [Mg/Fe] relations agree well with what limited direct
information we have concerning [Mg/Fe] in our DHG sample. Jablonka
et al. (\cite{Jablonkaetal1996}) did not include the bulge of the
\object{Milky Way} in their sample, but their prediction for
[Mg/Fe] is in excellent accord with McWilliam \& Rich's
(\cite{McWilliamRich1994}) [Mg/Fe] measurement.  The slope of
Jablonka et al.'s (\cite{Jablonkaetal1996}) relation is somewhat
steeper than the change in [O/Fe] we find between \object{M32} and
the bulge of the \object{Milky Way}, but easily within the error
bars.  J{\o}rgensen's (\cite{Jorgensen1997}) slope is shallower
than Jablonka et al.'s (\cite{Jablonkaetal1996}), but otherwise is
in excellent agreement with the change in [O/Fe] between
\object{M32} and the bulge of the \object{Milky Way}. The
behaviour of [Mg/Fe] and [O/Fe] with velocity dispersion indicates
that the fraction of the metallicity contributed by
$\alpha$-elements increases with increasing galaxy mass.  Given
our earlier arguments, this result implies a shorter time scale
for star formation in more massive ellipticals and bulges,
supporting the spectral synthesis conclusion that more massive
ellipticals are simpler than lower mass ellipticals from the
perspective of their star formation history (e.g., Worthey
\cite{Worthey1996}).

If a relation between [Mg/Fe] and velocity dispersion holds for
bulges and ellipticals, it implies an underlying relation between
the star formation time scale and the gravitational potential. The
data, however, imply that the time scale varies in a manner
opposite to that predicted in the classical model of galactic
winds (Larson \cite{Larson1974}), wherein winds are initiated
earliest in the lowest mass galaxies. Matteucci
(\cite{Matteucci1994}) and Tantalo et al.
(\cite{Tantaloetal1996}), among others, have successfully modelled
such \lq\lq inverse" winds by postulating an increasing efficiency
of star formation in more massive systems.  One advantage of a
longer time scale for star formation in lower luminosity DHGs is
that it would allow greater dissipation of the baryonic matter,
which might explain their higher central stellar densities.

Dwarf spheroidals clearly deviate from the [O/Fe] trend
established by bulges and ellipticals. Despite their low
luminosities and velocity dispersions, their [O/Fe] ratios
indicate that they formed the majority of their stars on a short
time scale, consistent with enrichment from type II supernovae
only, regardless of luminosity.  Nevertheless, dwarf spheroidals
did undergo chemical evolution, forming multiple generations of
stars, for more massive systems have higher oxygen abundances.
Dwarf spheroidals do fit the Larson (\cite{Larson1974}) picture in
the sense that a galactic wind was initiated very early in the
evolution of these low mass galaxies. However, within this galaxy
class, we cannot tell whether galactic winds were initiated later
in more massive systems, because they have uniformly high [O/Fe]
ratios.

\subsection{The abundance-luminosity relation}

We now turn our attention to the relationship between oxygen
abundances and luminosities in DHGs.  Figure \ref{fig5} compares
the oxygen abundances for the galaxies in Table \ref{tab1} with
oxygen abundances in dwarf irregular galaxies (Richer \& McCall
\cite{RicherMcCall1995}). The oxygen abundances for the DHGs are
the mean values for their stars.  In contrast, the oxygen
abundances for the dwarf irregulars are those for \ion{H}{ii}
regions, and therefore represent the oxygen abundances in the
interstellar medium, the highest abundance found in the stellar
component, not the mean. To the extent that dwarf irregulars share
a common star formation history, in a statistical sense at least,
the correction to mean stellar abundance would shift the
[O/H]-M$_B$ relation shown to lower abundances.

Figure \ref{fig5} demonstrates that all DHGs share a common
relationship between luminosity and oxygen abundance.
Statistically, the relationships for dwarf spheroidals and other
DHGs do not differ in either zero point or slope.  One immediate
implication is that the history of star formation, and its time
scale in particular, has no bearing upon the oxygen abundance that
is attained in DHGs.  This suggests that the energy reservoir
responsible for generating the galactic wind must store the energy
input from supernovae rather efficiently.

The common [O/H]-M$_B$ relation followed by all DHGs is
unexpected.  As is well-known, the luminosities of DHGs depend
upon both the surface brightness and the size of the system (e.g.,
Bender et al. \cite{Benderetal1992}), so a \lq\lq second
parameter" is apparently missing from Fig. \ref{fig5}.  However,
in the supernova-driven winds scenario, the forces that govern the
chemical evolution also govern the luminosity that is attained.
Chemical evolution ceases when a galactic wind is initiated.  The
luminosity in long-lived stars is also determined at this point,
for no further star formation may occur.  In this case, the
luminosity and oxygen abundance are functions of the same
fundamental parameters: the energy injection, the total mass, and
the mass distribution. Consequently, a \lq\lq second parameter"
may not be necessary in the [O/H]-M$_B$ relation.

This need not imply that all DHGs follow the same chemical
evolution.  In forming a given luminosity of long-lived stars, a
fixed mass of oxygen is injected into the interstellar medium
(supposing the IMF is constant). Injecting a particular mass of
oxygen need not yield a unique oxygen abundance, for the oxygen
abundance depends upon the mass of gas into which the oxygen is
injected and upon the importance of gas flows while star formation
takes place.  It is easy to imagine that the importance of gas
flows and the ease with which they can be driven depends upon the
shape or gradient of the potential. Thus, gas flows and gas
consumption could easily have had different relative importances
to the evolution of different classes of DHGs, if their mass
distributions were different.  Consequently, dwarf spheroidals and
ellipticals of similar luminosity need not convert the same
fraction of their original matter into stars, yet may nevertheless
attain similar oxygen abundances.

\section{Calibration of the Mg$_2$ index for ellipticals/bulges}

We can use the oxygen abundances in Table \ref{tab1} to make an
observational calibration of the Mg$_2$ index.  There are several
reasons why a calibration in terms of oxygen abundance is
reasonable.  Salaris et al. (\cite{Salarisetal1993}) found that
the total metal content and the combined abundance of C, N, O, and
Ne had similar affects upon the locations in the H-R diagram of
the main sequence and the main sequence turn-off.  Worthey
(\cite{Worthey1994}) has shown that main sequence and main
sequence turn-off stars are significant contributors to the light
in the blue and visual parts of the spectrum of old stellar
systems (contributing $> 45\%$ of the light).  Since the
abundances of $\alpha$-elements scale with the abundance of oxygen
and account for approximately 70\% of the total metallicity in the
sun, it is reasonable to expect the oxygen abundance and the
Mg$_2$ index to vary together.

Recently, Grillmair et al. (\cite{Grillmairetal1996}) determined a
mean age of 8.5\,Gyr for the stars in a field near the centre of
\object{M32}.  Although no comparable study of stars in  the inner
bulge of \object{M31} or the Milky Way exists, we can use the
recent globular cluster age re-calibration based upon Hipparcos
data (Reid \cite{Reid1997}) to set an upper limit of 13\,Gyr on
the age of their stars.  Concerning the Mg$_2$ index values, Table
\ref{tab1} provides nuclear values for M32 and the bulges of M31
and the Milky Way. Since we are not using nuclear velocity
dispersions (Sect. 3), we chose to compute the Mg$_2$ index values
appropriate to our adopted velocity dispersions using the Mg$_2 -
\sigma$ relation from Bender et al. (\cite{Benderetal1993}).

Based upon a multiple least-squares fit to the data, the predicted
Mg$_2$ index values vary with the oxygen abundance and age
according to
\begin{eqnarray}
\log\mathrm{Mg}_2 = (0.331 \pm 0.050)([\mathrm{O}/\mathrm{H}]+
    \log Age) \nonumber \\
    + (-0.857 \pm 0.037) \ \ \ \pm 0.022\,\mathrm{dex}\,.
\label{eqn5}
\end{eqnarray}
If the sensitivity of [O/H] is forced to be 1.68 times greater
than the sensitivity to age, as suggested by the studies of
Worthey (\cite{Worthey1994}) and Casuso et al.
(\cite{Casusoetal1996}), then
\begin{eqnarray}
\log\mathrm{Mg}_2 = (0.228 \pm 0.041)(1.68 [\mathrm{O}/\mathrm{H}] +
   \log Age) \nonumber \\
   + (-0.730 \pm 0.024) \ \ \ \pm 0.026\,\mathrm{dex}\,.
\label{eqn6}
\end{eqnarray}
In Eqs. \ref{eqn5} and \ref{eqn6}, Mg$_2$ is the Mg$_2$ index
measured in magnitudes, [O/H] is the logarithmic oxygen abundance
relative to solar (from Table \ref{tab1}), and $Age$ is the mean
stellar age measured in Gyr.  The quoted uncertainties are the
standard errors of the fits.  The relative age-oxygen abundance
sensitivity adopted in Eq. \ref{eqn6} should be valid provided the
fraction of the total metallicity represented by oxygen does not
depart too significantly from the solar value.

\section{Conclusions}

We have used the oxygen abundances in the planetary nebulae in
\object{M32}, the bulges of \object{M31} and the \object{Milky
Way}, and the dwarf spheroidals \object{NGC 205}, \object{NGC
185}, \object{Sagittarius}, and \object{Fornax} to derive mean
oxygen abundances for their stars.  The mean stellar oxygen
abundances are found to be modest, and would be easily attainable
in simple models of galactic evolution.  Combining the oxygen
abundances with the best estimates of the mean stellar ages, we
have calibrated the Mg$_2$ index as a function of age and oxygen
abundance.

The oxygen abundances in DHGs paint a very interesting picture of
chemical evolution.  Our principal result is that the oxygen
abundances in all DHGs correlate with their velocity dispersions.
Since the oxygen abundance is a measure of the energy input from
type II supernovae and the velocity dispersion is a measure of the
gravitational potential, a correlation between the two implies a
correlation between the energy input from supernovae and the
gravitational potential energy of the interstellar medium.  The
correlation improves if we account for the energy input from type
Ia supernovae.  A connection between energy input and the
gravitational potential arises naturally if a galactic wind is the
instrument that terminates chemical evolution.

The [O/Fe] ratios we derive from our oxygen abundances for
\object{M32} and the bulge of the \object{Milky Way} concur with
previous evidence from metallicity indicator studies that the
[$\alpha$-element/Fe] ratio increases with increasing luminosity
for bulges and ellipticals (Worthey et al. \cite{Wortheyetal1992};
Jablonka et al. \cite{Jablonkaetal1996}; J{\o}rgensen
\cite{Jorgensen1997}). Thus, the gravitational potential not only
determines the metallicity that is attained, but also fixes the
star formation time scale.  The sense of these
[$\alpha$-element/Fe] trends is such that more massive galaxies
develop galactic winds sooner than less massive galaxies, contrary
to the classical picture of galactic winds (Larson
\cite{Larson1974}).  Dwarf spheroidals do not participate in this
trend, and instead have [O/Fe] ratios indicative of uniformly
short time scales for star formation, regardless of luminosity.

Finally, we find that all DHGs follow a single relationship
between oxygen abundance and luminosity.  We argue that this
relation arises because both oxygen abundance and luminosity are
set by the interaction between star formation and the
gravitational potential.  This is a fundamental consequence of the
supernova-driven winds model of galaxy evolution. In forming a
given mass of stars, all DHGs achieve similar oxygen abundances,
but they need not follow the same chemical evolution.  The
systematically larger [O/Fe] ratios observed in dwarf spheroidals
indicate that the history of star formation in a DHG does not
affect the mean stellar oxygen abundance it attains.

\begin{acknowledgements}
MGR began this work during an extended visit in the Department of
Physics and Astronomy at York University, and so would like to
thank Marshall McCall for financial support and the Department of
Physics and Astronomy at York University for its warm
hospitality.  MLM thanks the Natural Sciences and Engineering
Research Council of Canada for its continuing support.
\end{acknowledgements}

\end{document}